\documentclass[superscriptaddress,twocolumn,prc,nofootinbib,showpacs]{revtex4}

\usepackage{graphicx}

\usepackage{amssymb}

\newcommand{\be}{\begin{equation}}
\newcommand{\ee}{\end{equation}}
\newcommand{\iso}[2]{{\ensuremath{{}^{#2}\mathrm{#1}}}}
\newcommand{\ovb}{$0\nu\beta\beta$}
\newcommand{\zvb}{$2\nu\beta\beta$}
\newcommand{\bbp}{$\beta^+\beta^+$}
\newcommand{\bbm}{$\beta^-\beta^-$}
\newcommand{\sussex}{Dept.~of Physics and Astronomy, University of Sussex, Brighton, BN1 9QH, UK}
\newcommand{\dort}{Lehrstuhl f\"ur Experimentelle Physik IV, Universit\"at Dortmund, Otto--Hahn Str.~4,44227 Dortmund, Germany}
\newcommand{\lngs}{Laboratori Nazionali del Gran Sasso, S.S. 17 BIS km. 18.910, 67010, Assergi, L'Aquila, Italy }
\newcommand{\bham}{School of Physics and Astronomy, University of Birmingham, B15 2TT, UK}
\newcommand{\liv}{Dept.~of Physics, University of Liverpool, Liverpool L69 7ZE, UK }
\newcommand{\war}{Dept.~of Physics, University of Warwick, Coventry CV4 7AL, UK}
\newcommand{\york}{Dept.~of Physics, University of York, Heslington, York, YO10 5DD, UK}
\begin{document}
\affiliation{\bham}
\affiliation{\dort}
\affiliation{\lngs}
\affiliation{\liv}
\affiliation{\sussex}
\affiliation{\war}
\affiliation{\york}
\flushbottom


\title{First Results on Double Beta Decay Modes of Cd, Te and Zn Isotopes with the COBRA
Experiment}
\author{T.~Bloxham}\affiliation{\bham}
\author{A.~Boston}\affiliation{\liv}
\author{J.~Dawson}\affiliation{\sussex}
\author{D. Dobos}\affiliation{\dort}
\author{S.P.~Fox}\affiliation{\york}
\author{M.~Freer}\affiliation{\bham}
\author{B.R.~Fulton}\affiliation{\york}
\author{C.~G\"o{\ss}ling}\affiliation{\dort}
\author{P.F.~Harrison}\affiliation{\war}
\author{M.~Junker}\affiliation{\lngs}
\author{H.~Kiel}\affiliation{\dort}
\author{J.~McGrath}\affiliation{\york}
\author{B.~Morgan}\affiliation{\war}
\author{D.~M\"unstermann}\affiliation{\dort}
\author{P.~Nolan}\affiliation{\liv}
\author{S.~Oehl}\affiliation{\dort}
\author{Y.~Ramachers}\affiliation{\war}
\author{C.~Reeve}\affiliation{\sussex}
\author{D.~Stewart}\affiliation{\war}
\author{R.~Wadsworth}\affiliation{\york}
\author{J.R.~Wilson}\affiliation{\sussex}
\author{K.~Zuber}\affiliation{\sussex}
\date{\today}

\begin{abstract}
\noindent
Four 1\,cm$^3$ CdZnTe semiconductor detectors were operated in the Gran Sasso
National Laboratory  to explore the feasibility of such devices for double beta
decay searches as proposed for the COBRA experiment. The research involved 
background studies accompanied by measurements of energy resolution performed
at the surface. Energy resolutions sufficient to reduce the contribution of
two-neutrino double beta decay events to a negligible level for a large scale
experiment have already been achieved and further improvements are expected.
Using activity measurements of contaminants in all construction materials a
background model was developed with the help of Monte Carlo simulations and
major background sources were identified. A total exposure of
4.34\,kg$\cdot$days of underground data has been accumulated allowing a search
for neutrinoless double beta decay modes of seven isotopes found in CdZnTe.
Half-life limits (90\% C.L.) are presented for decays to ground and excited
states. Four improved lower limits have been obtained, including zero neutrino
double electron capture transitions of \iso{Zn}{64}  and \iso{Te}{120} to the
ground state, which are $1.19\times10^{17}$\,years and 
$2.68\times10^{15}$\,years  respectively.  
\end{abstract}  
\maketitle
%


\section{Introduction}
\label{intro}
In recent years, a range of neutrino oscillation experiments~\cite{sk,sno,kaml}
have successfully proved that neutrinos are massive particles. Although such
experiments are sensitive to a mass-difference rather than absolute neutrino
mass, the data suggest a neutrino mass eigenstate of at least 50\,meV. To further
probe the neutrino's properties, it is necessary to look to other processes
such as neutrinoless double beta decay (\ovb), which violates lepton number by
two units; observation of this process would confirm the Majorana nature of the
neutrino and the rate of this rare decay is proportional to the absolute
neutrino mass scale. For recent reviews of double beta decay see
Refs.~\cite{review} and~\cite{review2}.\\
The COBRA experiment uses CdZnTe (CZT) semiconductors to search for \ovb
~\cite{zuber}. CZT contains nine double beta emitters, five of which can decay
via double  beta decay, {\it i.e.} emitting two electrons, and four of them via
either double electron capture, a combination of a positron emission with
electron capture or double positron emission. The study of the positron
and electron capture modes can be used for lepton number violating decay
searches on an equivalent level to \ovb, however the phase space for the
positron modes is strongly reduced, making them less sensitive. Nevertheless,
is has been shown that the positron/electron capture mixed modes have an
enhanced sensitivity to right-handed weak currents and thus can help to
disentangle the underlying physics mechanism of \ovb\ if
observed~\cite{hir94}. In addition, excited states transitions can be explored with 
high efficiency and low background using coincidence techniques among the detectors.
These decays would allow an independent search for double beta decay searching for the
de-excitation photons together with the electron signal.\\
The main focus of the work described in this paper is the study of background
through measurements performed underground, and energy resolution studies
carried out in a surface laboratory. Optimisation of these quantities is vital
for a successful search for \ovb, because, in the background limited case, the
observable half-life depends on them with a square root behaviour. In addition,
half-life limits for seven double beta isotopes contained in natural CZT have
been determined from data collected with a small prototype detector
accumulating an exposure of  4.34\,kg$\cdot$days. %
\section{Sensitivity}
\begin{figure}
  \centering
  \includegraphics[width=3.0in]{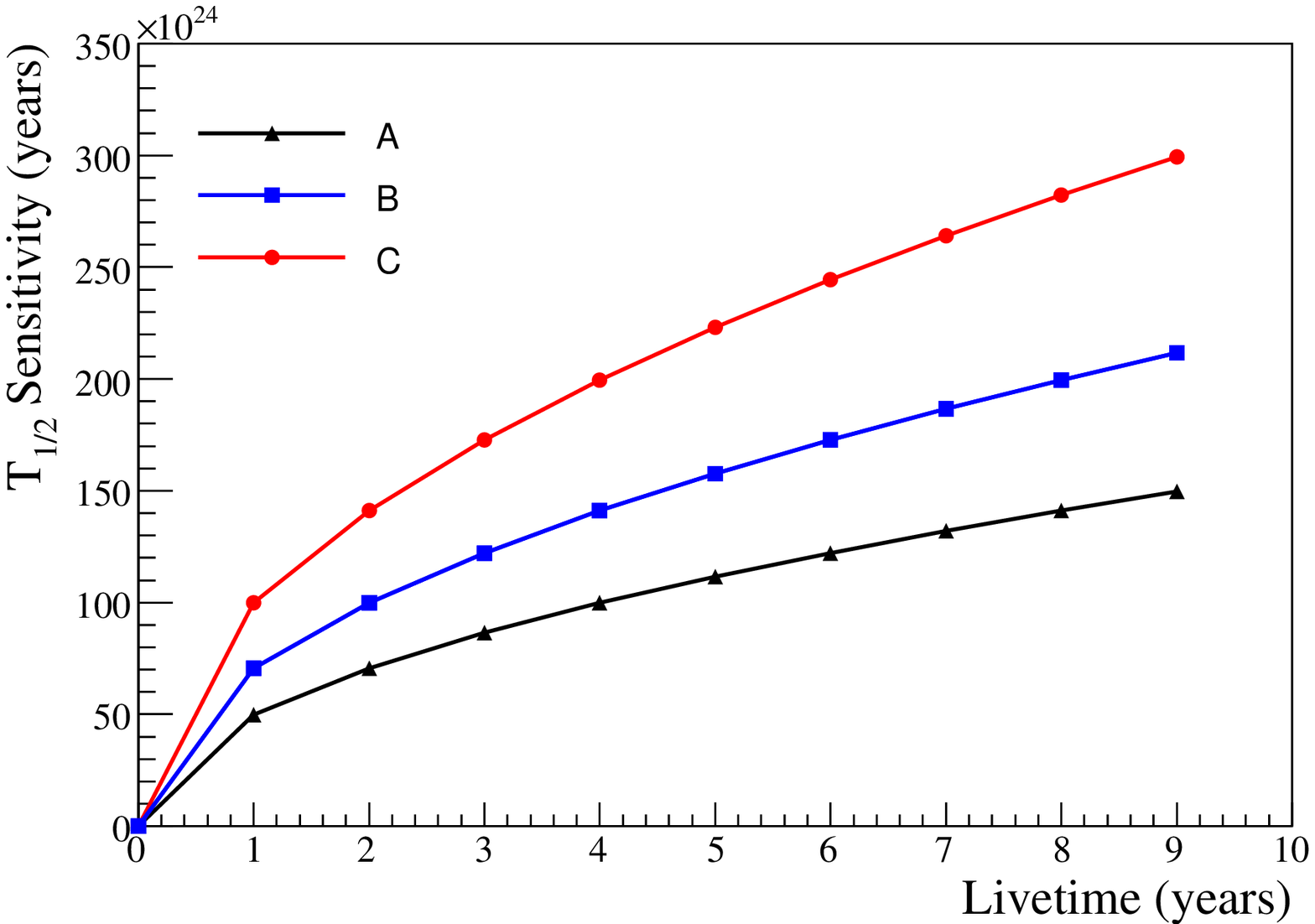}
  \caption{\label{f:sensi} (Color online) The expected half-life sensitivity for
$0\nu\beta\beta$--decay of \iso{Cd}{116} in an array of 64,000 1\,cm$^3$
detectors, enriched to 90\% in \iso{Cd}{116} for three scenarios of background
and energy resolution ($\Delta E$ = FWHM): \\ A =
$10^{-3}$ counts/(keV kg year), $\Delta E = 2\%$ at 2.8\,MeV. \\ B =
$10^{-3}$ counts/(keV kg year), $\Delta E = 1\%$ at 2.8\,MeV. \\ C =
$5\times10^{-4}$ counts/(keV kg year), $\Delta E = 1\%$ at 2.8\,MeV. 
A half life of about 2$\times10^{26}$ years corresponds to a neutrino mass
sensitivity of about 50 meV using matrix elements from Ref.~\cite{nme}.}
\end{figure}

Although COBRA is able to search for \ovb\ in a number of isotopes, the
sensitivity for the modes with lower Q-values will ultimately be limited by
background contributions from two neutrino double beta (\zvb) decays of the
isotopes with higher Q-values. The contribution of \zvb\ decay events to the
current data set is negligible, but for a very sensitive neutrino mass search,
COBRA will focus on \iso{Cd}{116}, which has the highest Q-value of 2809\,keV
for the nuclear decay to \iso{Pd}{116}. A peak will occur at this energy in
the sum energy spectrum for the case of \ovb.\\
Crucial experimental parameters, besides the mass of the detector/sample, are
energy resolution and the number of contaminating background events in this
range, as shown in Fig.~\ref{f:sensi}. Possible background sources include
cosmic rays, the natural radioactive decay chains, radioisotopes produced by
cosmic ray interactions within the materials used and neutrons. To get a first
glimpse of the background using CZT detectors, a prototype setup has been
installed in the Gran Sasso Underground Laboratory (LNGS) in Italy, which
provides an average shielding of $\sim$3500\,mwe against  cosmic ray sources. 
\section{Experimental Setup}
\label{s:setup}
The data presented here were obtained with four
1\,cm$\times$1\,cm$\times$1\,cm  CZT semiconductor detectors, each of mass
$\sim$6.5\,g, provided by eV-PRODUCTS. They utilise coplanar grid technology
to ensure that only the electron signal is read out~\cite{cpg} and hence
symmetric energy peaks are obtained.  All four crystals were operated at a
voltage of $-$1500\,V with a 20--40\,V grid bias applied between the two
anodes. The crystal surfaces, except for the gold-coated cathode side, are
covered with a passivation paint which prevents oxidation and deterioration in
detector performance over time.\\
The four detectors were mounted in a copper brick separated from all
preamplifier electronics by $\sim$25\,cm. The copper brick was part of a
$(20\,$cm)$^3$ cube of electro-polished copper which was embedded in a further
15\,cm of lead. The whole setup was located in a Faraday cage made from copper
plates. The cage was surrounded by a neutron shield, consisting of 7\,cm thick
boron-loaded polyethylene plates and an additional 20\,cm of paraffin wax at
the bottom.  This neutron shield was upgraded to cover 3 sides of the cage, as
well as the base, with paraffin wax before the third data taking period (period
C). Data collection commenced with a CAMAC based data acquisition (DAQ) system
in which the signals were fed into four peak sensing ADC modules (LeCroy 3511
and 3512) via custom-built preamplifiers and shaping main-amplifiers. Prior to
period C, the system was upgraded to a VME-based DAQ with four custom-built,
peak sensing, 14-bit ADC channels.
\section{Data
Acquisition}
\label{s:datasum}
\begin{figure}
  \centering
  \includegraphics[width=3.0in]{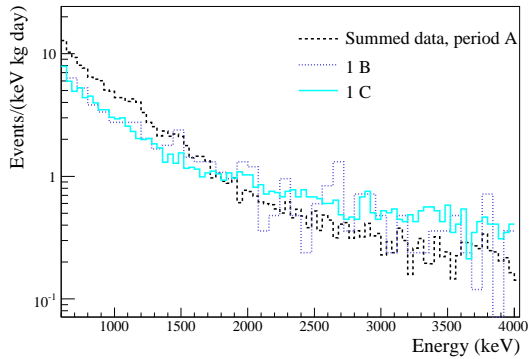}
  \caption{\label{f:dataplot}(Color online)  The prepared data sets for periods A, B and C.
  Period A comprises data from all 4 detectors and has slightly less events in 
  the high energy regions as the background levels in detectors 3 and 4 are 
  lower. However, periods B and C show less events at low energies due to
  improvements in the cleanliness of mounting materials.}
\end{figure}

The data analysed in this paper can be divided into three periods, separated by
upgrades to the experimental configuration that could have affected the
background contributions. In period A, the crystals were held in pertinax
mounting plates and connected directly to lemo cables. These mounting materials
were replaced by cleaner (in the radiopurity sense) delrin holders, whilst the
lemo connections were exchanged for  copper traces mounted on kapton foils at
the start of period B. Before period C the data acquisition hardware was
upgraded to the VME system and the paraffin neutron shielding was completed.\\ 
\begin{table}[t]
\begin{center}
\begin{tabular}{cccc}\hline\hline
Subset	& \multicolumn{2}{c}{Livetime (days)}	&Events/(keV kg day) \\ 
	& $>$500\,keV & $>$600\,keV & in 2--3\,MeV range  \\ \hline	
1A	& 32.86 & 150.99 & 0.57 $\pm$ 0.02\\ 
2A	& 0 	& 112.34 & 0.59 $\pm$ 0.03\\ 
3A	& 52.84	& 52.84  & 0.38 $\pm$ 0.03\\ 
4A	& 62.62	& 134.64 & 0.34 $\pm$ 0.02\\ 
1B	& 16.03	& 16.03  & 0.54 $\pm$ 0.07\\ 
1C	& 197.53& 197.53 & 0.56 $\pm$ 0.02\\ \hline\hline
\end{tabular}
\caption{\label{t:datasum} Livetimes for data sets prepared above two different 
energy thresholds (500\,keV and 600\,keV), and
an indication of the average background level in the 2--3\,MeV region.}
\end{center}
\end{table}
During the latter periods not all crystals were fully operational, so only data
from crystal 1 were analysed. In each period, individual runs were limited to
one hour and selection criteria were applied on a run-by-run basis to reject   
data affected by ``bursts'' of abnormally high event rates. Subsequent studies
have shown two main causes for such bursts: vibrations of the apparatus that
cause a piezo-electric effect in the crystals resulting in false event signals,
and breakdown effects due to faulty contacts to the crystal electrodes. To
reject the affected data, firstly the dead-time was calculated from the total
number of events per hour and the length of the event readout cycle during
which new events could be missed. 207 runs in which the dead-time exceeded 2\%
(due to large numbers of events with low ADC counts) were rejected. For the
remaining runs the distribution of the number of events per hour in the energy
range 300--4000\,keV was fitted with a Poissonian distribution. Runs with an
event rate exceeding the 99\% upper limit of the fitted distribution were
discarded from the analysis. A further 203 runs were rejected in this way. No
other cuts were applied to the data and in total 2.5\% of the runs were
rejected. \\
\noindent
It is possible that each crystal had different background contributions due to
intrinsic and surface contaminants (as indicated by the values in
table~\ref{t:datasum}). For this reason, data collected with each crystal, and
in each period, were considered as a different data sub-set. The summed data for
each of the collecting periods are shown in Fig.~\ref{f:dataplot}.\\
\noindent
Although detectors have been operated for extensive periods with thresholds 
below 100\,keV, at times it was necessary to raise the threshold for data
collection to exclude electronic phenomena. Therefore, to maximise the livetime
for analysis, two data sets have been prepared; one with an energy threshold of
600\,keV, and one with a 500\,keV threshold that omits any runs with thresholds
in the range 500--600\,keV. The livetimes for each subset and threshold are
given in table~\ref{t:datasum}. The total livetime for the high-threshold data
set is 4.34\,kg$\cdot$days, whilst the low-threshold data set comprises
2.36\,kg$\cdot$days.\\
\noindent
The energy resolution and stability of the detectors was calibrated regularly
with the help of \iso{Cs}{137}, \iso{Co}{60} and \iso{Th}{228} sources. 
Time-averaged resolution functions were determined for each data subset with 
all crystals showing a linear increase in FWHM with increasing energy. FWHM
values in the range 5--8\% at 2809\,keV were achieved.  Variations in the
resolution achieved can be attributed to changes in the  contacting methods and
the voltages applied between the different data taking periods.  It should be
noted that the detectors used here do not have the best energy resolution
possible, since for this first study with unknown background a very good energy
resolution was not considered to be essential and, hence, cheaper detectors
were used.

\section{Background studies }
\label{s:bg}
\begin{figure}
  \centering
  \includegraphics[width=3.0in]{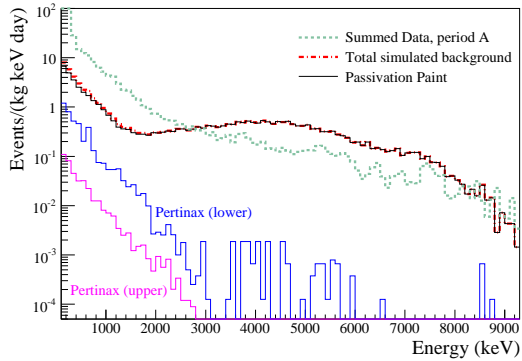}
  \caption{\label{f:bgmodel}(Color online) Summed data spectrum for  period A, 
  compared to simulated activities of the individual components. This clearly
  shows that events from the passivation paint on the detector surfaces
  contribute a significant background to the range of interest
  (2--3 MeV). The excess of events below 3\,MeV is thought to be due to
  backgrounds associated with the lemo cables and solder.}
\end{figure}

To understand the observed spectrum and disentangle the individual
contributions, a background model has been developed. All materials used were
measured for contaminants in the LNGS Ge-detector facility, though some of them
could only be measured after the start of data taking with the prototype. As a
consequence of these measurements, the pertinax holders and lemo cables were
replaced by delrin holders and kapton foils (between periods A and B). No
contamination of the CdZnTe could be detected with the Ge facility. With the
known activities of contaminants in the individual components, extensive Monte
Carlo modelling based on GEANT4 was performed to describe the observed spectrum
(Fig.~\ref{f:bgmodel}). \\
By far the largest background in the 2--3\,MeV region evolves from the
passivation paint on the detector surface. The precise prediction of this
contribution varies slightly due to the unknown paint mass and the
inhomogeneous paint thickness, which affects the alpha-particle simulation in
particular. However, there is a slight advantage associated with the
contaminated paint as the detector effectively acts as a self-calibrating
device. The installation of the VME system at the start of period C
significantly increased the timing resolution achieved for event read-out (from
$\approx$1\,ms to $\le$\,10\,$\mu$s) permitting the observation of
$\beta-\alpha$ coincidence events from \iso{Bi}{214}. This isotope originates
from the Th-decay chain, present in the passivation paint, and  beta-decays
with an endpoint of 3.3\,MeV; the daughter isotope, \iso{Po}{214}, alpha-decays
with a half-life of 164.3\,$\mu$s, releasing a 7.7\,MeV $\alpha$. The rate of
event pairs observed in the period C data-set was  consistent with  the
measured activity of a paint sample. Furthermore, it shows that any possible
dead layer at the detector surface is insignificant, otherwise the
alpha-particles would not be detected. In the meantime, an alternative solution
for surface passivation of the CZT detectors has been found and is currently
being explored at LNGS. Initial measurements show a reduction of this 
background by at least a factor of 8, if not more, in the region of interest
around 2.8\,MeV. \\
The measurements of pertinax contaminants prompted the replacement of all
pertinax components with delrin, leading to a reduction of events by about a
factor of 5 in the range 500--2000 keV, though some of this reduction can be
attributed to the exchange of the lemo cables.
\section{Achievable Energy resolution }
\label{s:eres}
The underground studies are not currently limited by energy resolution and,
therefore,  it was not considered necessary to use the highest quality of
crystals in this set-up. However, as background levels are reduced the energy
resolution will become important, since a sharp peak is especially important in
reducing the contribution of the irreducible background  of \zvb\ events to the
\ovb\ peak region. The fraction of \zvb\ events in the peak region, as a
function of energy resolution (FWHM) can be approximated  by~\cite{elliott}
\be
F = \frac{8Q}{m_e} \left(\frac{\Delta E}{Q}\right)^6
\ee
With this in mind, additional studies were performed outside the underground
experiment to determine the resolution achievable with CZT coplanar-grid
detectors and to investigate possible improvements. \par
\begin{figure}
  \centering
  \includegraphics[width=3.0in,height = 2.1in,angle=0]{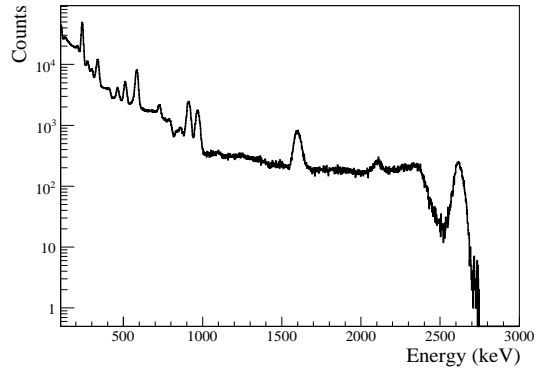}
  \caption{\label{f:thorium} Energy spectrum of a \iso{Th}{228} calibration
  source obtained with a 1\,cm$^3$ CZT  detector at room temperature of
  around 24 degrees. The energy resolution (FWHM) at the 2614 keV\,line is 
  2.6\%. }
\end{figure}

Fig.~\ref{f:thorium} shows a \iso{Th}{228} spectrum measured with a typical
`medium quality' CZT detector, resulting in a resolution of 2.6\% at 2614\,keV
but resolutions as good as 2.1\% have been measured with COBRA crystals.  With
such a resolution, for the case of 90\% \iso{Cd}{116} isotopic enrichment, 
\zvb\ decays will only contribute $2\times10^{-4}$ counts/(kg yr) to the
59\,keV wide signal region (calculated using the observed half-life of  $2.7
\times 10^{19}$\,yrs~\cite{sak06}). This is already well below the required
background levels shown in Fig.~\ref{f:sensi} but detectors with still better
energy resolution are commercially available and are being considered for use
in a future stage of the experiment. 
\\
A further experimental option to improve the energy resolution is cooling of
the detector. This might be especially important for searches in the low energy
range, for decays such as two neutrino double electron capture ($2\nu$ECEC)
that produce a signal below 100\,keV. First measurements of cooling from 
24$^\circ$C to 10$^\circ$C revealed an improvement in energy resolution of a
factor two below 100\,keV and an improvement of 5 \% on the typical resolution
at 2809\,keV. 
\section{Data Analysis}
\label{s:anal}
\begin{figure}
  \centering
  \includegraphics[width=3.0in,angle=0]{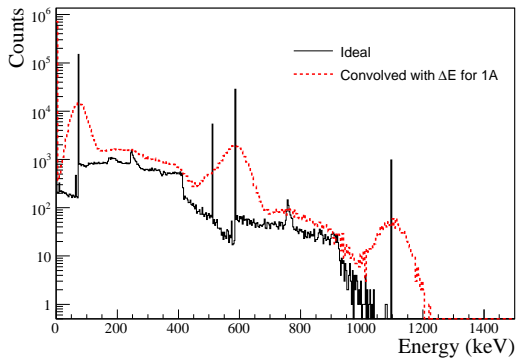}
  \caption{\label{f:zn64bpECgs}(Color online) Simulated spectrum in crystal 1
for $^{64}$Zn $\beta^+$EC decays to the ground state for ideal resolution and
convolved with the energy response for data subset 1A.  }
\end{figure}

The data analysis consists of two independent parts: simulation of
the possible double beta decays to determine detection efficiencies and a maximum
likelihood peak search.\\
The predicted signals in the crystals were determined through a GEANT4 based
Monte Carlo simulation utilising calculations from the Fortran Decay0
code~\cite{decay0}. \ovb\ decays to ground state (g.s.) and excited
states\footnote{Only excited states already included in the Decay0 code were
used, therefore some transitions, namely those to the higher excited states of
\iso{Te}{120}, have been omitted from this analysis.} were simulated for each
of the candidate isotopes contained in natural CdZnTe. For \bbm\ transitions,
calculations based on the light Majorana neutrino exchange mechanism were used
for ($0^+\rightarrow0^+$) transitions, whilst right-handed currents were used
in the calculation for ($0^+\rightarrow2^+$) transitions. As there is no general
connection between ground state and excited state matrix elements, they must be explored
separately for each isotope.\\
The energy, $E_{\rm peak}$, and intensity of the dominant peak for each \bbm\ 
decay were determined from these simulations and are given in 
table~\ref{t:params-minus}. The efficiency for observation of the full peak
energy, $\epsilon$, determined from the peak intensity generally decreases with
increasing peak energy. For decays to excited states, gamma escape
probabilities also play a part. \\
\begin{table}
\begin{center}
\begin{tabular}{rl|c|c|c|c}\hline
\hline
Isotope         & \hspace{0.3cm}Decay         & $E_{\rm peak}$ & $\Delta E_{\rm peak}$ & $\epsilon$ & Fit range \\ 
                &               & (MeV) & (\%) & (\%) & (MeV) \\ \hline
$^{116}$Cd      & to  g.s       	& 2.809 & 4.7--7.6 & 66.5 & 2.21--3.20	\\ 
$^{130}$Te      & to  g.s       	& 2.529 & 4.8--7.8 & 70.9 & 2.21--3.20	\\ \hline
$^{130}$Te      & to  536\,keV  	& 1.993 & 5.0--8.5 & 61.2 & 1.70--2.28	\\ \hline
$^{116}$Cd      & to  1294\,keV 	& 1.511 & 5.4--9.4 & 74.4 & 1.20--1.78	\\ \hline
$^{116}$Cd      & to  1757\,keV 	& 1.048 & 6.0--11.0 & 60.4 & 0.90--1.30	\\ \hline
$^{70}$Zn       & to  g.s       	& 1.001 & 6.1--11.3 & 93.3 & 0.60--1.30	\\ 
$^{128}$Te      & to  g.s       	& 0.868 & 6.5--12.2 & 94.8 & 0.60--1.30	\\ \hline
$^{116}$Cd      & to  2027\,keV 	& 0.778 & 6.8--12.9 & 67.4 & 0.50--1.20	\\ \hline
$^{116}$Cd      & to  2112\,keV 	& 0.693 & 7.1--13.8 & 77.3 & 0.50--1.00	\\ \hline
$^{116}$Cd      & to  2225\,keV 	& 0.580 & 7.7--15.4 & 76.6 & 0.50--1.00	\\ \hline
\hline
\end{tabular}
\caption{\label{t:params-minus}Specifications of the fits performed for $\beta^-\beta^-$ peaks. Values given for
each decay are the peak energy, $E_{\rm peak}$, the FWHM ($\Delta E_{\rm peak}$) at that peak energy (range for
all 6 data subsets), the efficiency, $\epsilon$, for observing that peak determined from simulations and the
energy range over which the fit was performed. Note that decays not separated by a horizontal line were fit
together.}
\end{center}
\end{table}
\noindent
For the isotopes that decay through \bbp\ transitions, double electron capture
(EC/EC), single electron capture ($\beta^+$EC) and double positron
(2$\beta^+$)-transitions were considered when energetically possible. In general
the predicted spectra for these decays are significantly more complex than those
for \bbm\ transitions, without clearly dominating peaks. As an example, the
simulated spectrum for $^{64}$Zn $\beta^+$EC decays to the ground state is shown
in Fig. \ref{f:zn64bpECgs}, both for ideal resolution and convolved with the
energy response for data subset 1A. Although there is a peak at the Q-value for
this decay (1096\,keV), there are also single and double 511\,keV escape peaks,
and a peak at 511\,keV due to gammas produced from the anihillation of
positrons produced by these decays in the other three crystals. \\
Detailed simulations and ex-situ measurements of background contaminants were
used to characterise the measured background continuum, as described in section 
\ref{s:bg}. 
An exponential of the form $y=A+Be^{-Cx}$ was found to describe the data well
at higher energies. Below 500\,keV the fits were less satisfactory due to the
presence of a number of low-energy gamma peaks and the 4-fold forbidden decay
of \iso{Cd}{113}~\cite{cd113} so a 500\,keV  threshold was enforced. Thus,
decay modes with no significant peaks above 500\,keV, namely those of
\iso{Cd}{108} and \iso{Cd}{114}, were omitted from the analysis presented in
this paper.\\
A study of the residuals of the background fits showed some evidence for
additional peaks in the continuum at 610\,keV and 1120\,keV, relating to gamma
lines from \iso{Bi}{214} decay. Due to the combined effects of decay branching
ratios, energy resolution and reduced efficiency for stopping higher energy
gammas in a single crystal, no other background gamma peaks within the analysis
region were predicted through simulations. Therefore, two gaussian peaks were
added to the description of the background with fixed mean and width determined
from the relevant resolution function. The amplitudes of these peaks were
treated as additional fit parameters. As a cross-check, fits for the
exponential background were performed without these additional background peaks
and in all cases resulted in a poorer, or negligibly different, chisquared
probability.\\
A maximum likelihood fit was performed to determine the most likely number of
signal events, $\theta_s$, over the combined data-set. Parameters describing
the background were allowed to vary between crystals and data collection
periods, but the \ovb\  signal rate was assumed to be constant throughout. 
{\it i.e.} Different background parameters were applied to different data
subsets to allow for the varying background rates (indicated in 
table~\ref{t:datasum}) but the normalised background distributions fitted to
each data subset were found to agree within errors for each fit scenario.\\
For \bbm\ modes, $\theta_s$ enters the fit through the amplitude of a gaussian
peak with width determined by the calibrated resolution of the relevant data
subset. The range of peak widths (FWHM) for each fitted peak are given in
table~\ref{t:params-minus} along with the energy range used for each fit.
Simulations showed that a range of $(E_{\rm peak} \pm 3 \Delta E)$ or greater
was required for each peak-search in order to adequately characterise the
background continuum. The close proximity of some predicted signal peaks
required these ground-state transition signals to be determined simultaneously:
$^{116}$Cd and $^{130}$Te were fit together and the $^{70}$Zn and $^{128}$Te
peaks were also fit simultaneously. Transitions to excited states are expected
to be significantly suppressed with respect to ground state transitions due to
phase-space arguments, so any contribution of excited state decays to the
fitted peaks for ground-state transitions is assumed to be negligible. The
limit for each signal arising from a transition to an excited state was
determined in a separate fit. The high-threshold ($>600$\,keV) data set was
used for all \bbm-mode peak searches except the decays to the third and fourth
excited states of \iso{Cd}{116}.\\
The majority of spectra predicted for \bbp-mode decays have multiple peaks, 
each significantly smaller in amplitude than those arising from \bbm\ decays to
the ground state, thus justifying separate treatment in the analysis.  Due to
the complexity of these spectra, to determine the most likely number of signal
events, $\theta_s$, the most likely scaling factor for the entire simulated
spectrum, was extracted from the likelihood fit. For each decay the simulated
spectrum was normalised to unity and convolved with the relevant resolution
function for each data sub-set. The range for each fit (as given in
table~\ref{t:results-plus}) was selected to include all the dominant peaks. For
transitions where the simulated spectra showed a dominant peak in the region
500--600\,keV, the low-threshold data set was used; in all other cases,  the
high-threshold data set was used. \\  
For all \bbm\ and \bbp\ modes analysed, a 90\% confidence limit on the
half-life, $T_{\rm half}$, was determined from the fitted number of signal
events, $\theta_s \pm \delta_s$, where $\delta_s$ is one sigma uncertainty in
the fit, for each decay under investigation.
\begin{equation}
T_{\rm half} = \frac{N_{\rm iso} t_{\rm live} \epsilon \ln2 }{\left(\theta_s +
1.28\delta_s\right)}
\label{e:thalflim}
\end{equation}
Here $N_{\rm iso}$ is the number of candidate nuclei per crystal for the  given
decay, $t_{\rm live}$ is the total duration of data collection in
crystal-years, and $\epsilon$ is an efficiency factor determined from
simulations. For \bbm-searches, $\epsilon$ is the fraction of simulated events
in the peak region (see table~\ref{t:params-minus}). However, for \bbp-searches
$\epsilon=1$ since the whole simulated spectrum, normalised to unity, is used
in the fit. \\
For each fit, a chisquared goodness of fit test was performed. However, due to 
the low statistics, this parameter was not expected to follow a true chisquared
distribution. Therefore, the distribution of the $\chi^2$ statistic was
determined by Monte Carlo for each fit in order to calculate the fit
probability. The $\chi^2$, and its respective probability, determined for each
\bbm-fit is included in table~\ref{t:results-minus}. For the \bbp-fits the
probabilities for each fit were also all $>$80\%. As a cross-check, fits were
repeated with $\theta_s$ fixed to zero for each signal, all of which resulted
in either negligible change or a decrease in the goodness of fit.\\
A detailed study of possible systematic effects was performed and the dominant
uncertainties were found to be those that affect the number of candidate
nuclei. Uncertainties in energy resolution and livetime, and possible biases in
the fit procedure, were all found to have a negligible effect on the analysis.
However, the possible existence of a dead-layer at the surface of the crystals
could reduce the active volume by up to 10\%. Observations of
$\alpha$-particles from the passivation paint indicate that the dead layer is
probably smaller than this but the effect was taken into account in a
conservative manner by using $N_{\rm iso}\times0.9$ in the limit calculation.
Due to the production process, the zinc content is only known to be in the
range 7--11\% resulting in an uncertainty in both the number of zinc nuclei and
the number of cadmium nuclei. To ensure conservative half-life limits, 7\% zinc
content was used when calculating the number of zinc nuclei, and 11\%  zinc
content was used in calculations for cadmium isotopes.

\section{Results}
\label{s:results}
\begin{figure}
  \centering
\begin{tabular}{c}  
       \includegraphics[width=3.0in]{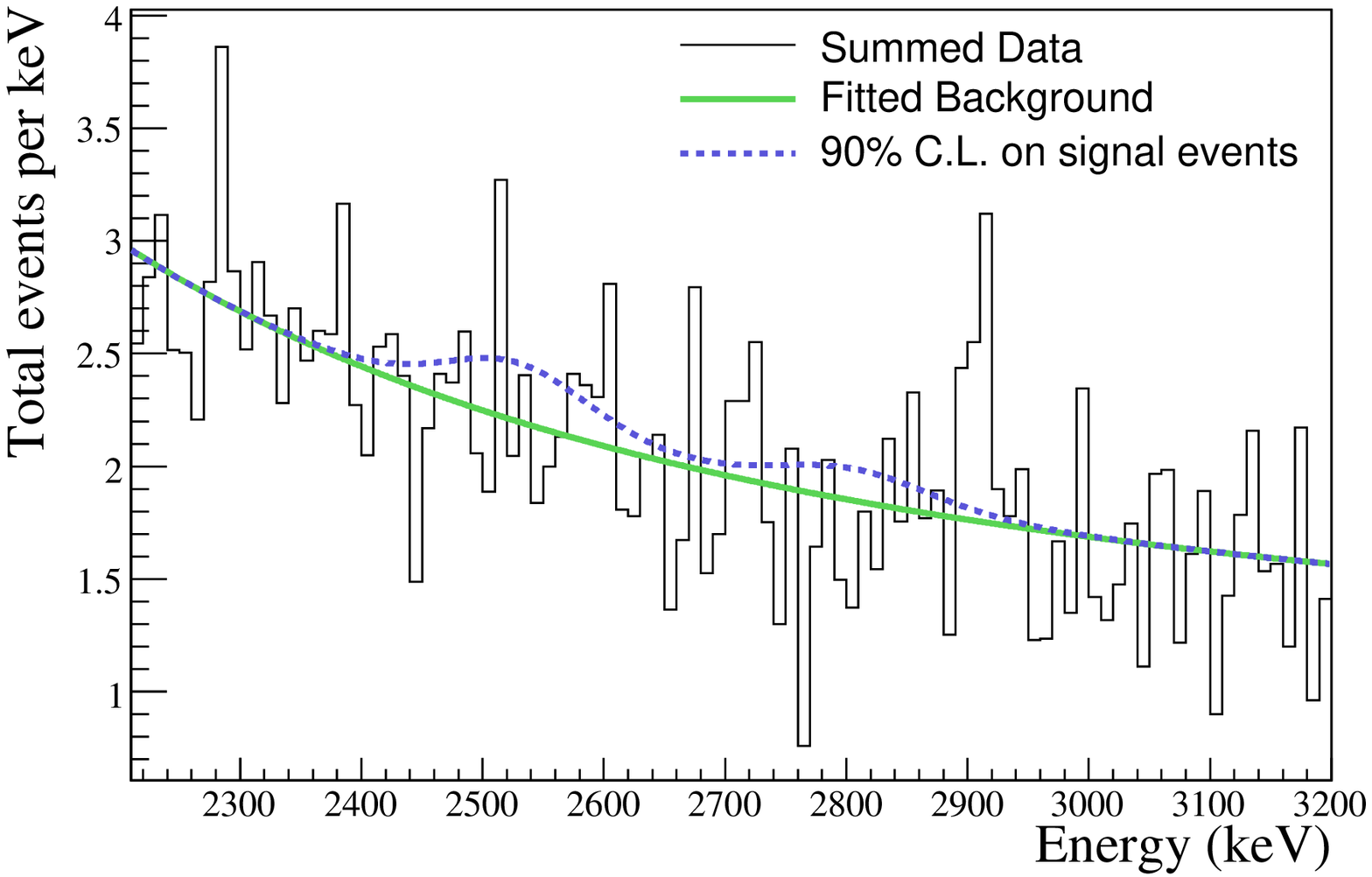} \\
       \includegraphics[width=3.0in]{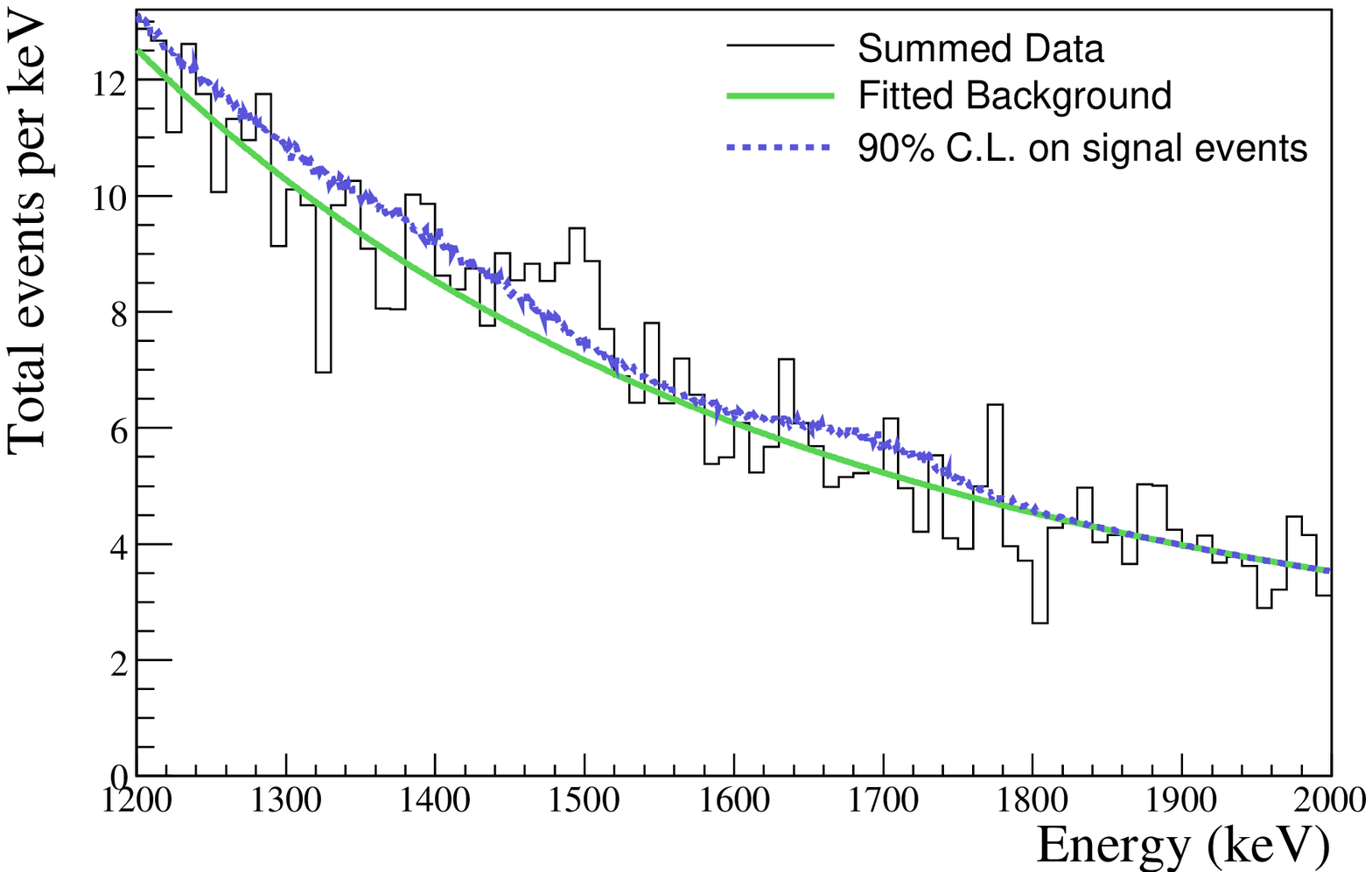}
\end{tabular}
       \caption{\label{f:fit}(color online) The fitted 90\% confidence limits
       for \iso{Cd}{116} and \iso{Te}{130} \ovb\ decay events (top) and 
       \iso{Te}{120} $0\nu$EC/EC to ground state events (bottom) with
       the total high-threshold data set shown for the range of each fit. Also
       shown separately is the contribution from the fitted background.}
\end{figure}

Table~\ref{t:results-minus} shows all the \bbm\-decay half-life limits (90\% C.L.) 
calculated in this work and table~\ref{t:results-plus} shows the limits
calculated for \bbp\-decays. The limits obtained from the combined fit to
the \iso{Cd}{116} and \iso{Te}{130} \ovb\ decays to ground state and for the 
\iso{Te}{120} double electron capture decay to ground state are shown in
Fig.~\ref{f:fit}.  
Due to the small detector mass, at present the searches for $^{116}$Cd and 
$^{130}$Te cannot compete with other running large scale experiments.
However, half-life limits obtained for \iso{Zn}{64} and \iso{Te}{120}
improve on existing measurements.
\begin{table}
\begin{center}
\begin{tabular}{rl|c|c|c|c}\hline
 \multicolumn{2}{c|}{Isotope and Decay}        & $\chi^2$/DoF& P &  \multicolumn{2}{|c}{T$_{1/2}$ limit (years)}   \\ 
                &               &  &  &             This work & World Best  \\ \hline \hline
$^{116}$Cd      & to  g.s       & 1236/2950 & 0.99 &3.14$\times 10^{19}$    &1.7$\times 10^{23}$~\cite{dan03}\\ 
$^{130}$Te      & to  g.s       & 1236/2950 & 0.99 &9.92$\times 10^{19}$    &1.8$\times 10^{24}$~\cite{cuoricino} \\ \hline
$^{130}$Te      & to  536\,keV   & 726/1703 & 0.99 &3.73$\times 10^{19}$    &9.7$\times 10^{22}$~\cite{alessandrello00} \\ \hline
$^{116}$Cd      & to  1294\,keV & 676/1703 & 1.00 &4.92$\times 10^{18}$    &2.9$\times 10^{22}$~\cite{dan03} \\ \hline
$^{116}$Cd      & to  1757\,keV & 615/1479 & 0.99 &9.13$\times 10^{18}$    &1.4$\times 10^{22}$ ~\cite{dan03} \\ \hline
$^{70}$Zn       & to  g.s       & 859/2078 & 1.00 &2.24$\times 10^{17}$   &9.0$\times 10^{17}$~\cite{znwo4} \\ 
$^{128}$Te      & to  g.s       & 859/2078 & 1.00 &5.38$\times 10^{19}$   &1.1$\times 10^{23}$~\cite{arnaboldi02}\\ \hline
$^{116}$Cd      & to  2027\,keV & 756/1732 & 0.96 &1.37$\times 10^{19}$    &2.1$\times 10^{21}$~\cite{pie94}\\ \hline
$^{116}$Cd      & to  2112\,keV & 545/1233 & 0.87 &1.08$\times 10^{19}$    &6.0$\times 10^{21}$~\cite{dan03}\\ \hline
$^{116}$Cd      & to  2225\,keV & 545/1233 & 0.87 &9.46$\times 10^{18}$    &1.0$\times 10^{20}$$^\dag$~\cite{barabash90} \\ \hline
\hline
\end{tabular}
\caption{\label{t:results-minus} Results from fits for $\beta^-\beta^-$decay
peaks.  The $\chi^2$/DoF goodness of fit parameter, and respective probability
determined through Monte Carlo are included.  The 90\% confidence limits have  
conservative systematic uncertainties applied and are compared to the world best
limits. $^\dag$Quoted limit is 68\% not 90\%. }
\end{center}
\end{table}

\begin{table}
\begin{center}
\begin{tabular}{rl|c|c|c}\hline
\multicolumn{2}{c|}{Isotope and Decay}          & Range&  \multicolumn{2}{|c}{T$_{1/2}$ limit (years)}   \\ \hline
                &               & (MeV)&              This work & World Best  \\ \hline \hline
$^{64}$Zn &$0\nu\beta^+$EC to g.s.      &0.5--1.3 & 2.78$\times$10$^{17}$    & 2.4$\times 10^{18}$~\cite{znwo4} \\ \hline
$^{64}$Zn &$0\nu$2EC to g.s.           &0.7--1.3 & {\bf 1.19$\times$10$^{17}$}    & 7.0$\times 10^{16}$~\cite{znwo4} \\ \hline
$^{120}$Te&$0\nu\beta^+$EC to g.s.      &0.5--2.0 & {\bf 1.21$\times$10$^{17}$}    & 2.2$\times 10^{16}$~\cite{kiel03}\\ \hline
$^{120}$Te&$0\nu$2EC to g.s.           &1.2--2.0 & {\bf 2.68$\times$10$^{15}$}    & -\\ \hline
$^{120}$Te&$0\nu$2EC to 1171keV        &0.5--2.0 & {\bf 9.72$\times$10$^{15}$}    & -\\ \hline
$^{106}$Cd&$0\nu\beta^+\beta^+$ to g.s. &0.5--2.0 & 4.50$\times 10^{17}$   & 2.4$\times 10^{20}$~\cite{belli99}\\ \hline
$^{106}$Cd&$0\nu\beta^+$EC to g.s.      &1.4--3.0 & 7.31$\times 10^{18}$   & 3.7$\times 10^{20}$~\cite{belli99}\\ \hline
$^{106}$Cd&$0\nu$2EC to g.s.           &1.4--3.0 & 5.70$\times 10^{16}$   & 1.5$\times 10^{17}$~\cite{norman84}\\ \hline
$^{106}$Cd&$0\nu\beta^+\beta^+$ to 512keV &0.5--2.0 & 1.81$\times 10^{17}$ & 1.6$\times 10^{20}$~\cite{belli99}\\ \hline
$^{106}$Cd&$0\nu\beta^+$EC  to 512keV   &0.8--2.0 & 9.86$\times 10^{17}$   & 2.6$\times 10^{20}$~\cite{belli99}\\ \hline
\hline
\end{tabular}
\caption{\label{t:results-plus} 90\% confidence limits obtained for all $\beta^+\beta^+$~decays
analysed in this work with conservative systematic uncertainties applied,
compared to the world best limits. New world best values from this work are
shown in bold. The energy range used for each fit is also included.}
\end{center}
\end{table}

\section{Summary}
\label{s:sum}
A new double beta decay experiment, COBRA, is planned using a large amount of
CZT semiconductor detectors. A low rate of background events in the peak region
and good energy resolution are crucial aspects of the design of such an
experiment. To develop this new approach, CZT semiconductor detectors have 
been operated deep underground, for the first time, to study their background.
Using a small prototype, a background model has been developed and a major
background component in the form of a passivation paint on the detector
surfaces was identified. Alternatives are now available and major improvements
are expected soon. \\
Studies of the attainable energy resolution showed that the contribution of
the  irreducible background of \zvb\ can be kept to a negligible level.  A
4.34\,kg$\cdot$day data set, collected with four 1 cm$^3$ crystals, has been
analysed to determine limits on various neutrinoless double beta decay modes of
seven different isotopes. Despite the small detector mass, only 26\,g of
CZT, these data have yielded four improved half-life limits for decays of
\iso{Zn}{64} and \iso{Te}{120}. After submission of this, paper new limits for
\iso{Te}{120} have been presented~\cite{bar07}.\\ 
In the near future, 64 CZT detectors will be running and improvements on all
the limits presented in this paper can be expected. In addition to the
increased detector mass, new criteria based on coincident energy deposits in
time and space will allow better rejection of backgrounds, and a full
characterisation of each individual crystal will help reduce systematic
uncertainties.

\section{Acknowledgements}
This research was supported by PPARC and the Deutsche Forschungsgemeinschaft (DFG).
We thank G.~Cowan for useful discussions, V.~Tretyak for providing the
Decay0 code and eV-PRODUCTS for their support. 
In addition, we thank the Forschungszentrum Karlsruhe, especially
K.~Eitel, for providing the material for the neutron shield. We thank the
mechanical workshop of the University Dortmund for their support and the
Laboratori Nazionali del Gran Sasso (LNGS) for offering the possibility to
perform measurements underground. The work has been supported by the TA-DUSL activity of the ILIAS
program (Contract No. RII3-CT-2004-506222) as part of the EU FP6
programme.

\bibliography{firstcobra}

\begin{thebibliography}{24}
\expandafter\ifx\csname natexlab\endcsname\relax\def\natexlab#1{#1}\fi
\expandafter\ifx\csname bibnamefont\endcsname\relax
  \def\bibnamefont#1{#1}\fi
\expandafter\ifx\csname bibfnamefont\endcsname\relax
  \def\bibfnamefont#1{#1}\fi
\expandafter\ifx\csname citenamefont\endcsname\relax
  \def\citenamefont#1{#1}\fi
\expandafter\ifx\csname url\endcsname\relax
  \def\url#1{\texttt{#1}}\fi
\expandafter\ifx\csname urlprefix\endcsname\relax\def\urlprefix{URL }\fi
\providecommand{\bibinfo}[2]{#2}
\providecommand{\eprint}[2][]{\url{#2}}

\bibitem[{\citenamefont{{Fukuda} et~al.}()}]{sk}
\bibinfo{author}{\bibfnamefont{Y.}~\bibnamefont{{Fukuda}}}
  \bibnamefont{et~al.}, \bibinfo{note}{\protect{Super-Kamiokande}
  Collaboration, Phys. Rev. Lett., {\bf 81}, 1562 (1998); {\bf 82}, 1810
  (1999); {\bf 82}, 2430 (1999); {\bf 86}, 5651 (2001)}.

\bibitem[{\citenamefont{{Ahmad} et~al.}()}]{sno}
\bibinfo{author}{\bibfnamefont{Q.}~\bibnamefont{{Ahmad}}} \bibnamefont{et~al.},
  \bibinfo{note}{\protect{SNO} Collaboration, Phys. Rev. Lett., {\bf 87},
  071301 (2001); {\bf 89}, 011301 (2002); {\bf 89}, 011302 (2002)}.

\bibitem[{\citenamefont{{Eguchi} et~al.}()}]{kaml}
\bibinfo{author}{\bibfnamefont{K.}~\bibnamefont{{Eguchi}}}
  \bibnamefont{et~al.}, \bibinfo{note}{\protect{KamLAND} Collaboration, Phys.
  Rev. Lett., {\bf 90}, 021802 (2003); {\bf 94} 081801 (2005)}.

\bibitem[{\citenamefont{{Zuber}}(2006)}]{review}
\bibinfo{author}{\bibfnamefont{K.}~\bibnamefont{{Zuber}}},
  \bibinfo{journal}{Acta Polonica} \textbf{\bibinfo{volume}{B37}},
  \bibinfo{pages}{1905} (\bibinfo{year}{2006}).

\bibitem[{\citenamefont{{Elliott} and {Engel}}(2004)}]{review2}
\bibinfo{author}{\bibfnamefont{S.}~\bibnamefont{{Elliott}}} \bibnamefont{and}
  \bibinfo{author}{\bibfnamefont{J.}~\bibnamefont{{Engel}}},
  \bibinfo{journal}{J. Phys.} \textbf{\bibinfo{volume}{G30}},
  \bibinfo{pages}{R183} (\bibinfo{year}{2004}).

\bibitem[{\citenamefont{{Zuber}}(2001)}]{zuber}
\bibinfo{author}{\bibfnamefont{K.}~\bibnamefont{{Zuber}}},
  \bibinfo{journal}{Phys. Lett.} \textbf{\bibinfo{volume}{B519}},
  \bibinfo{pages}{1} (\bibinfo{year}{2001}).

\bibitem[{\citenamefont{{Hirsch} et~al.}(1994)}]{hir94}
\bibinfo{author}{\bibfnamefont{.}~\bibnamefont{{Hirsch}}} \bibnamefont{et~al.},
  \bibinfo{journal}{Z. f. Phys.} \textbf{\bibinfo{volume}{A347}},
  \bibinfo{pages}{151} (\bibinfo{year}{1994}).

\bibitem[{\citenamefont{{Rodin} et~al.}(2006)}]{nme}
\bibinfo{author}{\bibfnamefont{V.}~\bibnamefont{{Rodin}}} \bibnamefont{et~al.},
  \bibinfo{journal}{Nucl. Phys.} \textbf{\bibinfo{volume}{A766}},
  \bibinfo{pages}{107} (\bibinfo{year}{2006}).

\bibitem[{\citenamefont{{Luke}}(1995)}]{cpg}
\bibinfo{author}{\bibfnamefont{P.}~\bibnamefont{{Luke}}},
  \bibinfo{journal}{IEEE Transactions on Nuclear Science}
  \textbf{\bibinfo{volume}{42}}, \bibinfo{pages}{207} (\bibinfo{year}{1995}).

\bibitem[{\citenamefont{{Elliott} and {Vogel}}(2002)}]{elliott}
\bibinfo{author}{\bibfnamefont{S.}~\bibnamefont{{Elliott}}} \bibnamefont{and}
  \bibinfo{author}{\bibfnamefont{P.}~\bibnamefont{{Vogel}}},
  \bibinfo{journal}{Ann. Rev. Nucl. Part. Sci.} \textbf{\bibinfo{volume}{52}},
  \bibinfo{pages}{115} (\bibinfo{year}{2002}).

\bibitem[{\citenamefont{{Saakyan}}(2006)}]{sak06}
\bibinfo{author}{\bibfnamefont{R.}~\bibnamefont{{Saakyan}}},
  \bibinfo{journal}{Talk at ILIAS Double beta Meeting, Valencia}
  (\bibinfo{year}{2006}).

\bibitem[{\citenamefont{{Ponkratenko} et~al.}(2000)\citenamefont{{Ponkratenko},
  {Tretyak}, and {Zdesenko}}}]{decay0}
\bibinfo{author}{\bibfnamefont{O.~A.} \bibnamefont{{Ponkratenko}}},
  \bibinfo{author}{\bibfnamefont{V.}~\bibnamefont{{Tretyak}}},
  \bibnamefont{and}
  \bibinfo{author}{\bibfnamefont{Y.}~\bibnamefont{{Zdesenko}}},
  \bibinfo{journal}{Phys. Atom. Nucl.} \textbf{\bibinfo{volume}{63}},
  \bibinfo{pages}{1282} (\bibinfo{year}{2000}), \bibinfo{note}{{\it The actual
  generator used was Decay0, not Decay4}}.

\bibitem[{\citenamefont{{Goessling} et~al.}(2005)}]{cd113}
\bibinfo{author}{\bibfnamefont{C.}~\bibnamefont{{Goessling}}}
  \bibnamefont{et~al.}, \bibinfo{journal}{Phys. Rev.}
  \textbf{\bibinfo{volume}{C72}}, \bibinfo{pages}{064328}
  (\bibinfo{year}{2005}).

\bibitem[{\citenamefont{{Danevich} et~al.}(2003)}]{dan03}
\bibinfo{author}{\bibfnamefont{F.~A.} \bibnamefont{{Danevich}}}
  \bibnamefont{et~al.}, \bibinfo{journal}{Phys. Rev.}
  \textbf{\bibinfo{volume}{C68}}, \bibinfo{pages}{035501}
  (\bibinfo{year}{2003}).

\bibitem[{\citenamefont{{Arnaboldi} et~al.}(2005)}]{cuoricino}
\bibinfo{author}{\bibfnamefont{C.}~\bibnamefont{{Arnaboldi}}}
  \bibnamefont{et~al.}, \bibinfo{journal}{Phys. Rev. Lett.}
  \textbf{\bibinfo{volume}{95}}, \bibinfo{pages}{142501}
  (\bibinfo{year}{2005}).

\bibitem[{\citenamefont{{Alessandrello} et~al.}(2000)}]{alessandrello00}
\bibinfo{author}{\bibfnamefont{A.}~\bibnamefont{{Alessandrello}}}
  \bibnamefont{et~al.}, \bibinfo{journal}{Phys. Lett.}
  \textbf{\bibinfo{volume}{B486}}, \bibinfo{pages}{13} (\bibinfo{year}{2000}).

\bibitem[{\citenamefont{{Danevich} et~al.}(2005)}]{znwo4}
\bibinfo{author}{\bibfnamefont{F.~A.} \bibnamefont{{Danevich}}}
  \bibnamefont{et~al.}, \bibinfo{journal}{Nucl. Instr. Meth.}
  \textbf{\bibinfo{volume}{A544}}, \bibinfo{pages}{553} (\bibinfo{year}{2005}).

\bibitem[{\citenamefont{{Arnaboldi} et~al.}(2003)}]{arnaboldi02}
\bibinfo{author}{\bibfnamefont{C.}~\bibnamefont{{Arnaboldi}}}
  \bibnamefont{et~al.}, \bibinfo{journal}{Phys. Lett.}
  \textbf{\bibinfo{volume}{B557}}, \bibinfo{pages}{167} (\bibinfo{year}{2003}).

\bibitem[{\citenamefont{{Piepke} et~al.}(1994)}]{pie94}
\bibinfo{author}{\bibfnamefont{A.}~\bibnamefont{{Piepke}}}
  \bibnamefont{et~al.}, \bibinfo{journal}{Nuclear Physics A}
  \textbf{\bibinfo{volume}{577}}, \bibinfo{pages}{493} (\bibinfo{year}{1994}).

\bibitem[{\citenamefont{{Barabash} et~al.}(1990)\citenamefont{{Barabash},
  {Kopylov}, and {Cherehovsky}}}]{barabash90}
\bibinfo{author}{\bibfnamefont{A.}~\bibnamefont{{Barabash}}},
  \bibinfo{author}{\bibfnamefont{A.}~\bibnamefont{{Kopylov}}},
  \bibnamefont{and}
  \bibinfo{author}{\bibfnamefont{V.}~\bibnamefont{{Cherehovsky}}},
  \bibinfo{journal}{Phys. Lett. B} \textbf{\bibinfo{volume}{249}},
  \bibinfo{pages}{186} (\bibinfo{year}{1990}).

\bibitem[{\citenamefont{Kiel et~al.}(2003)\citenamefont{Kiel, M\"unstermann,
  and Zuber}}]{kiel03}
\bibinfo{author}{\bibfnamefont{H.}~\bibnamefont{Kiel}},
  \bibinfo{author}{\bibfnamefont{D.}~\bibnamefont{M\"unstermann}},
  \bibnamefont{and} \bibinfo{author}{\bibfnamefont{K.}~\bibnamefont{Zuber}},
  \bibinfo{journal}{Nucl. Phys.} \textbf{\bibinfo{volume}{A723}},
  \bibinfo{pages}{499} (\bibinfo{year}{2003}).

\bibitem[{\citenamefont{{Belli} et~al.}(1999)}]{belli99}
\bibinfo{author}{\bibfnamefont{P.}~\bibnamefont{{Belli}}} \bibnamefont{et~al.},
  \bibinfo{journal}{Astropart. Phys.} \textbf{\bibinfo{volume}{10}},
  \bibinfo{pages}{115} (\bibinfo{year}{1999}).

\bibitem[{\citenamefont{{Norman} and {DeFaccio}}(1984)}]{norman84}
\bibinfo{author}{\bibfnamefont{E.}~\bibnamefont{{Norman}}} \bibnamefont{and}
  \bibinfo{author}{\bibfnamefont{M.}~\bibnamefont{{DeFaccio}}},
  \bibinfo{journal}{Phys. Lett.} \textbf{\bibinfo{volume}{B148}},
  \bibinfo{pages}{31} (\bibinfo{year}{1984}).

\bibitem[{\citenamefont{{Barabash} et~al.}(2007)}]{bar07}
\bibinfo{author}{\bibfnamefont{A.}~\bibnamefont{{Barabash}}}
  \bibnamefont{et~al.}, \bibinfo{journal}{Preprint}
  \textbf{\bibinfo{volume}{nucl-ex/0703020}} (\bibinfo{year}{2007}).

\end{thebibliography}
\end{document}